\documentclass{amsart}
\usepackage{amssymb, amsmath}

\newtheorem{thm}{Theorem}[section]
\newtheorem{lemma}[thm]{Lemma}

\theoremstyle{definition}

\newcommand{\beq}{\begin{equation}}
\newcommand{\eeq}{\end{equation}}
\newcommand{\bea}{\begin{eqnarray}}
\newcommand{\eea}{\end{eqnarray}}

\renewcommand{\Im}{\operatorname{\rm Im}\nolimits}

\def \comp {\operatorname{comp}}

\def \restrict {\upharpoonright}
\def \Real {{\mathbb R}}

\def \Complex {\mathbb{C}}
\def \Natural {{\mathbb N}}

\title [The resonance counting function]
{The Resonance Counting Function
for Schr\"odinger Operators with Generic Potentials}

   \author { T.\ Christiansen and P.\ D.\ Hislop}
\thanks{TC partially supported by NSF grant DMS 0088922 and PDH partially
supported by NSF grant DMS 0202656.}
\begin{document}

\begin{abstract}
We show that the resonance counting function for a Schr\"odinger operator
has maximal order of growth for generic sets of real-valued, or
complex-valued,
$L^\infty$-compactly
supported potentials.
\end{abstract}
\maketitle
\section{Introduction}

The purpose of this note is to show that for a generic set of
compactly supported potentials,
the resonance counting function for the associated
Schr\"odinger operator has maximal order of growth.
We consider odd dimensions $d \geq 1$, and any potential
$V\in L^{\infty}_{\comp}(\Real^d)$. We
define the set of {\it scattering poles} or
{\it resonances} of the Schr\"odinger operator
$H_V \equiv - \Delta + V$ on $L^2 ( \Real^d)$ through the meromorphic
continuation of the resolvent. To make this precise,
let $\chi_V$ be a smooth, compactly
supported function equal to one on the support of $V$.
It is well-known that the operator-valued
function $\lambda \rightarrow \chi_V (H_V - \lambda^2)^{-1} \chi_V$ admits
a meromorphic continuation (denoted by the same symbol)
from $\Im \lambda \geq 0$, taken as the
physical half-plane, to the entire complex plane. The poles of this
continuation (including multiplicities)
are independent of the choice of $\chi_V$ satisfying these
conditions.  There are
at most a finite number of poles with $\Im \lambda > 0$
corresponding to the finitely-many eigenvalues
of $H_V$.
The set of scattering poles of $H_V$ is
defined by
\bea
\label{resonance1}
\mathcal{R}_V & = &\{ \lambda_j\in \Complex:\; \chi_V (H_V-\lambda^2)^{-1}
\chi_V    \\
& &   \text{has a pole at }\; \lambda=\lambda_j,\; \text{listed with
multiplicity}\} .  \nonumber
\eea
This definition can be made for both
real-valued and complex-valued potentials.
The {\it resonance counting function} $N_V(r)$ for $H_V$ on $L^2 (
\Real^d)$, is defined as
\beq
\label{defn1}
N_V(r)=\#\{\lambda_j\in \mathcal{R}_V: \; |\lambda_j|<r\}.
\eeq

The large $r$ properties of $N_V (r)$ have been extensively
studied,
and we refer the reader to the review article of Zworski \cite{[Zworski1]}.
The leading asymptotic behavior is known in one dimension
\cite{[Froese1],[Simon1],[Zworski2]},
and for certain spherically symmetric potentials
for odd $d \geq 3$ \cite{zwrp}.
Moreover, the following upper bound on
$N_V(r)$ for compactly supported potentials
is well-known
\beq
\label{upperbd1}
N_V (r) \leq C_{V,d} ( 1 + r^d ) ,
\eeq
see, for example, \cite{[Froese],[Melrose1],[Vodev],{zwrp}, {zwodd}}.
In addition, for nontrivial
real-valued, compactly supported potentials,
it is known that an infinite number of
resonances exist \cite{[Melrose2],[SaBZ2]}.
More recently,
S\'a Barreto \cite{[SaB]} proved a lower bound of the form
\beq
\label{lowerbd1}
\lim\sup_{r \rightarrow \infty} \frac{N_V(r)}{r} > 0,
\eeq
for nontrivial $V \in C_c^\infty ( \Real^d; \Real)$.
The situation is different for complex-valued,
$L^\infty$ compactly supported
potentials.
There are nontrivial examples of
such potentials {\it with no resonances} for $d \geq 3$
\cite{[Christiansen1]}.

The purpose of this note is to prove that the resonance counting
function $N_V (r)$, defined in (\ref{defn1}), has the maximal order
of growth $d$ for
a generic family of
either real-, or complex-valued,
compactly supported potentials.
Following B.\ Simon \cite{simon}, for a metric space $X$, we call a dense
$G_{\delta}$ set $S\subset X$ {\it Baire typical}.
Our main result is the following
theorem.

\begin{thm}\label{thm:main}
Let $d\geq 3 $ be odd, let $K\subset \Real^d$ be a compact set with
nonempty interior, and let $F=\Real$ or $F=\Complex$.  Then the set
$$
\mathcal{M}=\{ V\in L^{\infty}(K;F): \lim\sup_{r\rightarrow \infty}
\frac{\log N_V(r)}{\log r} =d\}
$$
is Baire typical in $L^{\infty}(K;F)$.
\end{thm}

The term {\it generic} is often used to
designate a property that typically occurs
for a given family. If $X$ is a complete metric space,
a property is {\it generic}
if it holds for a family $\mathcal{F}$ of dense $G_\delta$ sets in $X$.
Such a family is closed under countable intersections and
has the property that if $A \in \mathcal{F}$, and $X$ is perfect, then $A
\cap
B_X$ is uncountable for any open ball $B_X \subset X$.
In this sense, our theorem says that the
resonance counting function for a generic family of
real- or complex-valued, $L^\infty$ compactly supported
potentials has the maximum order of growth
given by the dimension $d \geq 1$.
Since there are nontrivial,
complex-valued, $L^\infty$ compactly supported potentials
for which $N_V (r)$ has  zero order of growth \cite{[Christiansen1]},
and since $N_0 (r)$ for the Laplacian (zero real potential)
has zero order of growth, our result is the best possible.
We remark that it would be interesting
to find nontrivial
potentials $V \in L_{\comp}^\infty ( \Real^d; \Real)$, $d \geq 3$, for
which the order of growth of $N_V (r)$ is {\it strictly less than $d$}.


\section{Proof of Theorem \ref{thm:main}}

We shall denote the scattering matrix for $H_V = - \Delta +V$
by $S_V(\lambda)$.  The operator $S_V(\lambda)$ acts on $L^2(S^{d-1})$
and if
$V$ is real-valued, then it is a unitary operator for $\lambda \in \Real$.
The $S$-matrix is given explicitly by
\begin{equation}
\label{eq:sm}
S_V(\lambda)=I+c_d \lambda^{d-2} \pi_{\lambda}
(V-VR_V(\lambda)V)\pi^t_{-\lambda} \equiv I + T_\lambda,
\end{equation}
where $R_V(\lambda)=(H_V-\lambda^2)^{-1}$ and $(\pi_{\lambda}f)
(\omega)=\int e^{-i\lambda x\cdot \omega}f(x)dx$ \cite{[Yafaev]}.
Under the assumption that $V \in L_{\comp}^\infty ( \Real^d; F)$,
the operator $T_\lambda : L^2 ( S^{d-1}) \rightarrow L^2 ( S^{d-1} )$ is
trace class.
The $S$-matrix has a meromorphic continuation to the entire
complex plane with finitely many poles in $\Im \lambda  > 0$,
corresponding to eigenvalues
of $H_V$, and resonances in $\Im \lambda < 0$.
We recall that if $\Im \lambda_0 \geq 2\|V\|_{L^{\infty}}+1$,
the multiplicities of $\lambda_0$, as a zero of $\det S_V(\lambda)$, and
$-\lambda_0$, as a pole of $(H_V-\lambda^2)^{-1}$, coincide, cf.\
Section 3 of \cite{zwpf}.
We will work with the function $\det S_V(\lambda)$.
For $N,\;M,\;q>0$, $j>2N+1$, let
\begin{multline*}
A(N,M,q,j)= \{ V\in L^{\infty}(K;F):\|V\|_{L^{\infty}}\leq N,\;
\log |\det(S_V(\lambda))|\leq M |\lambda|^q  \\
\; \text{for}\; \Im \lambda \geq 2N+1 \; \text{and}\; |\lambda|\leq j\}.
\end{multline*}
We remark that $\det S_V(\lambda)$ is holomorphic in this region.

\begin{lemma}
\label{closed1}
The set $A(N,M,q,j) \subset L^\infty ( \Real^d)$ is closed.
\end{lemma}
\begin{proof}
Let $V_k\in A(N,M, q, j)$, such that $V_k\rightarrow V$ in the $L^{\infty}$
norm.
Then clearly $\|V\|_{L^\infty} \leq N$.
We shall use (\ref{eq:sm})
and the bound
\begin{equation}\label{eq:detbd}
|\det(I+A)-\det(I+B)|\leq \| A-B\|_1e^{\|A\|_1+\|B\|_1+1},
\end{equation}
cf.\ \cite{[Simon2]}. We let $\| \cdot \|_1$ and $\| \cdot \|_2$
denote the trace and Hilbert-Schmidt norms, respectively, on $L^2 (
S^{d-1})$.
We wish to show that $\|S_{V_k}(\lambda)-S_V(\lambda)\|_1 \rightarrow
0$ as $k\rightarrow \infty$. Let $\chi \in C_c^{\infty}(\Real^d)$
be a function that is equal to one on $K$.
Using (\ref{eq:sm}), we have
\begin{eqnarray*}
\lefteqn{ \|S_{V_k}(\lambda)-S_V(\lambda)\|_1  } \nonumber \\
& \leq &  |c_d||\lambda|^{d-2}\|\pi_{\lambda}\chi\|_2
(\|V_k-V\|_{L^{\infty}}+ \|V_kR_{V_k}V_k-VR_V V\|_{L^2\rightarrow L^2})
\|\chi \pi^t_{-\lambda}\|_2 .
\end{eqnarray*}
As in Lemma 3.3 of \cite{[Froese]}, using the explicit Schwartz kernel of
$\pi_{\lambda}$, one can see that if $|\lambda|\leq j$ there is a constant
$C_j$ such that $\|\pi_{\lambda}\chi\|_2\leq C_j$ and
$\|\chi \pi_{-\lambda}^t\|_2 \leq C_j$.
We need only show that
$\|V_kR_{V_k}V_k-VR_V V\|_{L^2\rightarrow L^2}\rightarrow 0$ as
$k\rightarrow
\infty$.  But since $\Im \lambda \geq 2N+1 \geq 2\max(\|V_k\|_{L^{\infty}},
\|V\|_{L^{\infty}})+1$, the operators $R_{V_k}(\lambda)$ and $R_V(\lambda)$
are
holomorphic functions of $\lambda$, with norms that are uniformly
bounded in this region.
Since
$$R_{V_k}(\lambda)-R_V(\lambda)= R_{V_k}(\lambda)(V-V_k)R_V(\lambda),$$
$\|R_{V_k}(\lambda)-R_V(\lambda)\|\rightarrow 0$ as $k\rightarrow \infty$.
Thus
$\|V_kR_{V_k}V_k-VR_V V \|_{L^2\rightarrow L^2}\rightarrow 0$ as
$k\rightarrow
\infty$.

A similar argument shows that $\|I-S_{V_k}(\lambda)\|_1$ and
$\|I-S_{V}(\lambda)\|_1$ are bounded uniformly for $\Im \lambda
\geq 2N+1$, $|\lambda|\leq j$.  Using (\ref{eq:detbd}), then, we
see that $\det S_{V_k}(\lambda)\rightarrow \det S_{V}(\lambda)$ and thus
$$\log |\det S_V(\lambda)| \leq M|\lambda|^q\; \text{if}\;
\Im \lambda \geq 2N+1 \; \text{and} \; |\lambda|\leq j.$$
\end{proof}

In the next step, we characterize
those $V \in L_{\comp}^\infty (K; F)$ for which the order of
growth of the resonance counting function
is strictly less than the dimension $d$.
For $N,\;M,\;q>0$, let
\begin{equation*}
B(N,M,q)=\bigcap_{j\geq 2N+1}A(N,M,q,j).
\end{equation*}
Note that $B(N,M,q)$ is closed by Lemma \ref{closed1}.

\begin{lemma}\label{l:inB}
Let $V\in L^{\infty}(K;F)$, with $$\lim \sup _{r\rightarrow
\infty}\frac{\log N_V(r)}{\log r}<d.$$  Then there exist $N,\; M,\; l\in
\Natural$ such that $V\in B(N,M,d-1/l)$.
\end{lemma}
\begin{proof}
By \cite[Lemma 4.2]{[Christiansen2]}, there is a
$p<d$ such that
$$\lim\sup_{r\rightarrow \infty}\frac{\log \max_{0<\theta<\pi}
\log|\det S_V(2 \|V\|_{L^{\infty}}+1+re^{i\theta})|}{\log r}
=p.$$  In fact, the continuity of
$\det S_V(\lambda)$ in this region implies that this bound is
true for $0\leq \theta\leq \pi$.
It follows that there is a $p'\geq p$, $p'<d$,
and an $M\in \Natural $ such that
$$ \log |\det S_V(\lambda)| \leq M |\lambda|^{p'}$$
when $\Im \lambda \geq 2\|V\|_{\infty}+1.$  Choose $l\in \Natural $
so that $p'\leq d-1/l$ and $N\in \Natural$ so that $N\geq \|V\|_{\infty}$,
and then  $V\in B(N,M,d-1/l)$ as desired.
\end{proof}

\begin{lemma}
\label{l:gd}
The set $$\mathcal{M}=\{ V\in L^{\infty}_{\comp}(K;F): \;
\lim \sup_{r\rightarrow \infty}\frac{\log N_V(r)}{\log r}=d\}$$
is a $G_{\delta}$ set.
\end{lemma}
\begin{proof} By Lemma \ref{l:inB}, the complement of
$\mathcal{M}$ is contained in
\begin{equation*}
\bigcup_{(N,M,l) \in \Natural^3}  B(N,M,d-1/l),
\end{equation*}
which is an $F_{\sigma}$ set since it is a countable union of
closed sets.  By \cite[Lemma 4.2]{[Christiansen2]}, if $V\in \mathcal{M},$
then $V\not \in B(N,M,d-1/l)$ for any $N,\; M,\; l \in \Natural.$
Thus $\mathcal{M}$ is the complement of an $F_{\sigma}$ set.
\end{proof}

\noindent
We can now prove our theorem.
\begin{proof}[Proof of Theorem \ref{thm:main}]
Since Lemma \ref{l:gd} shows that $\mathcal{M}$ is a $G_{\delta}$ set,
we need only show that $\mathcal{M}$
is dense in $L^{\infty}(K;F)$.  To do this,
we use a slight modification of the proof of \cite[Corollary
1.3]{[Christiansen2]}.
We give the proof here for the convenience of the reader.
Let $V_0\in L^{\infty}(K;F)$ and let $\epsilon >0$.  By \cite[Theorem
2]{zwrp}, we may choose a spherically symmetric $V_1\in
L^{\infty}(K;\Real)$ so that $V_1\in \mathcal{M}$ and $\|V_1\|_{L^{\infty}}
<\epsilon/2$. We now
consider the function $V(z) \equiv V(z,x)=zV_1(x)+(1-z)V_0(x)$.
This potential
satisfies the conditions of \cite[Theorem 1.1]{[Christiansen2]}, and
$V(0)=V_0$.
Thus, by \cite[Theorem 1.1]{[Christiansen2]},
for some pluripolar set $E\subset \Complex$,
we have
$$
\lim\sup_{r\rightarrow \infty}\frac{\log N_{V(z)}(r)}{\log r}=d,
$$
for $z\in \Complex
\setminus E$.  In particular, since $E \restrict \Real \subset \Real$ has
Lebesgue measure $0$
(e.g. \cite[Section 12.2]{ransford}),
we may choose a point $z_0\in \Real$, $z_0\not \in E$, with
$|z_0|<\epsilon/{2(1+\|V_0 \|_{L^{\infty}})}$.  Then $V(z_0)\in \mathcal{M}$
and $\|V(z_0)-V_0\|_{L^{\infty}}<\epsilon$.  Note that if $V_0$ is
real-valued (respectively, complex-valued) then so is $V(z_0)$.
\end{proof}


\small
\noindent
{\sc
Department of Mathematics\\
University of Missouri\\
Columbia, Missouri 65211\\
e-mail:{\tt tjc@math.missouri.edu} }

\vspace{2mm}

\noindent
{\sc Department of Mathematics\\
University of Kentucky\\
Lexington, Kentucky 40506--0027\\
e-mail: {\tt hislop@ms.uky.edu}}
\end{document}